\documentclass[runningheads]{svmult}

\usepackage{makeidx}   % allows index generation
\usepackage{graphicx}  % standard LaTeX graphics tool
\usepackage{subeqnar}  % subnumbers individual equations
\usepackage{multicol}  % used for the two-column index
\usepackage{physprbb}  % modified textarea for proceedings,
\makeindex             % used for the subject index
\newcommand{\etal}{$\it et\, al.$\,}
\begin{document}
\title*{The evolution of galaxy mass in hierarchical models}
\toctitle{The evolution of galaxy mass in hierarchical models}
% allows explicit linebreak for the table of content
%
%
\titlerunning{The evolution of galaxy mass}
% allows abbreviation of title, if the full title is too long
% to fit in the running head
%
\author{C. M. Baugh\inst{1}
\and A. J. Benson\inst{2}
\and S. Cole\inst{1}
\and C. S. Frenk\inst{1}
\and C. Lacey\inst{1}}
\authorrunning{Baugh et al.}
% if there are more than two authors,
% please abbreviate author list for running head
%
%
\institute{
Department of Physics, Durham University, South Road, Durham, DH1 3LE.
\and 
Caltech, MC105-24, 1200 E. California Blvd., Pasadena, CA 91125}

\maketitle              % typesets the title of the contribution

\begin{abstract}
Advances in extragalactic astronomy have prompted the development 
of increasingly realistic models which aim to describe the formation 
and evolution of galaxies. We review the philosophy behind one such 
technique, called semi-analytic modelling, and explain the relation 
between this approach and direct simulations of gas dynamics. 
Finally, we present model predictions for the evolution of 
the stellar mass of galaxies in a universe in which structure 
formation is hierarchical.
\end{abstract}

\section{Modelling the formation and evolution of galaxies}
An incredibly wide range of physical processes are believed 
to be influential in the formation of galaxies. 
Some of these processes are well understood, for example, the 
build up of dark matter haloes through mergers or the accretion 
of smaller units; the formation of haloes has been studied extensively  
using N-body simulations and can be described analytically with a 
reasonable degree of success (e.g. Lacey \& Cole 1993, 1994). 
On the other hand, we are still some distance away from being able to 
simulate the formation of stars. An impressive initial step in this 
direction has been taken by Abel \etal (2002) with a simulation 
that leads up to the formation of the first star in the universe. 
However, the conditions in this calculation 
are much simpler than would be typical for the formation of the bulk 
of the stars in the universe and the simulation is stopped once 
additional physics not currently included in the calculation, 
such as radiative transfer, 
become important.   

The absence of a complete theory of star formation need not be an 
obstacle to the development of a theory of galaxy formation. One can 
take a phenomenological approach in which a physically motivated 
recipe is adopted to describe star formation within a galaxy. 
The recipe will inevitably contain one or more uncertain parameters  
but these can be fixed by comparing model predictions with observational data. 

Adopting this pragmatic approach, two techniques have been 
developed to model the formation and evolution of galaxies.
The first of these is the direct simulation of gravitational instability 
and gas dynamics. 
The second class of technique is semi-analytic modelling (Kauffmann 
\etal 1993; Cole \etal 1994). In such models, the merger trees of dark 
matter haloes can either be grown using a Monte-Carlo algorithm or 
they can be extracted from an N-body simulation. The gas physics, namely 
shock heating, radiative cooling, star formation and supernova feedback 
(along with galaxy mergers), is followed using approximations and simple 
rules. 

The two techniques have complementary pros and cons. Direct 
simulations do not require the specialised assumptions that are 
necessary in the semi-analytic models, e.g. the 
imposition of spherical symmetry in the calculation of the gas cooling 
time. On the other hand, semi-analytic models are fast and flexible, 
allowing a wide range of parameter space to be explored. 
The modular structure of the semi-analytic models means that new 
prescriptions for processes such as star formation can be readily 
evaluated.

In certain respects, the two techniques are actually very similar. The direct 
simulation approach necessarily breaks down at some level because of the 
finite resolution that is attainable. It is not possible to achieve 
the sub-parsec resolution needed to simulate star formation in a 
cosmologically representative volume. Coupled with the lack 
of knowledge of the relevant micro-physics, this means that recipes 
like those used in the semi-analytic models have to be deployed in order 
to produce a fully specified model. 

The first comparisons of the two techniques have recently been 
carried out (Benson \etal 2001a; Helly \etal 2002 ; Yoshida \etal 2002). 
These studies considered the rate at which gas cools in SPH simulations 
and in ``stripped-down'' semi-analytic models in which star formation 
and feedback have been switched off. The two approaches are in remarkably 
good agreement, which inspires confidence in the cooling model adopted 
in the semi-analytic schemes.

\section{ Constructing a model}
\begin{figure}
\begin{center}
\includegraphics[width=1.02\textwidth]{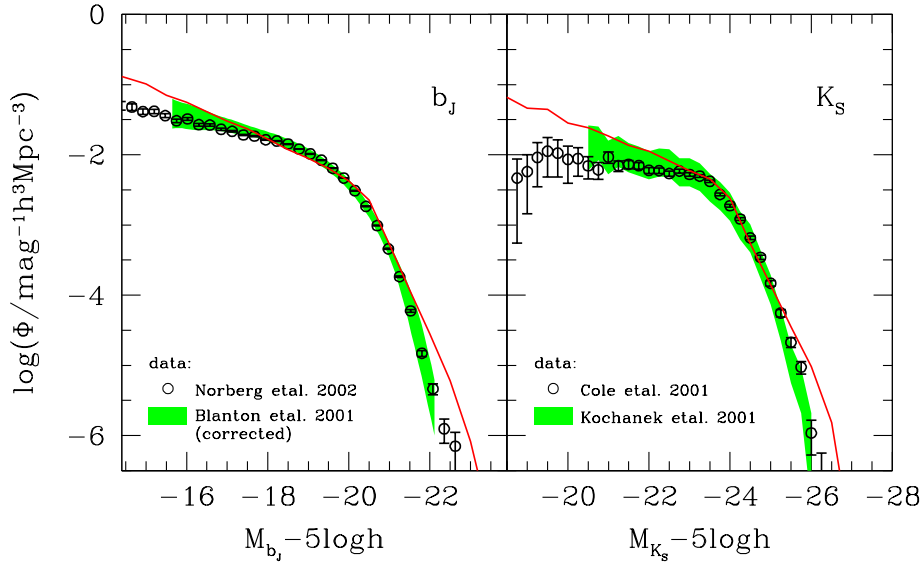}
\end{center}
\caption[]
{
The local galaxy luminosity function, in the $b_{J}$- and $K_{S}$- bands.
The predictions of the fiducial model from Cole \etal (2000) are shown by 
the solid line in each panel. In the left panel, the symbols shows an 
estimate of the luminosity function from the 2dFGRS (Norberg \etal 2002). 
The shaded region shows an estimate based on the analysis of SDSS data 
in Blanton \etal (2000) (see Norberg \etal 2002 for full details).
In the right panel, a combination of 2dFGRS redshifts and 2MASS photometry 
was used to estimate the near infrared luminosity function (Cole \etal 2001).
The shaded region shows another observational estimate which 
also uses 2MASS photometry (Kochanek \etal 2001).
}
\label{fig:lf}
\end{figure}

\begin{figure}
\begin{center}
\hspace{-1cm}
\includegraphics[width=0.8\textwidth]{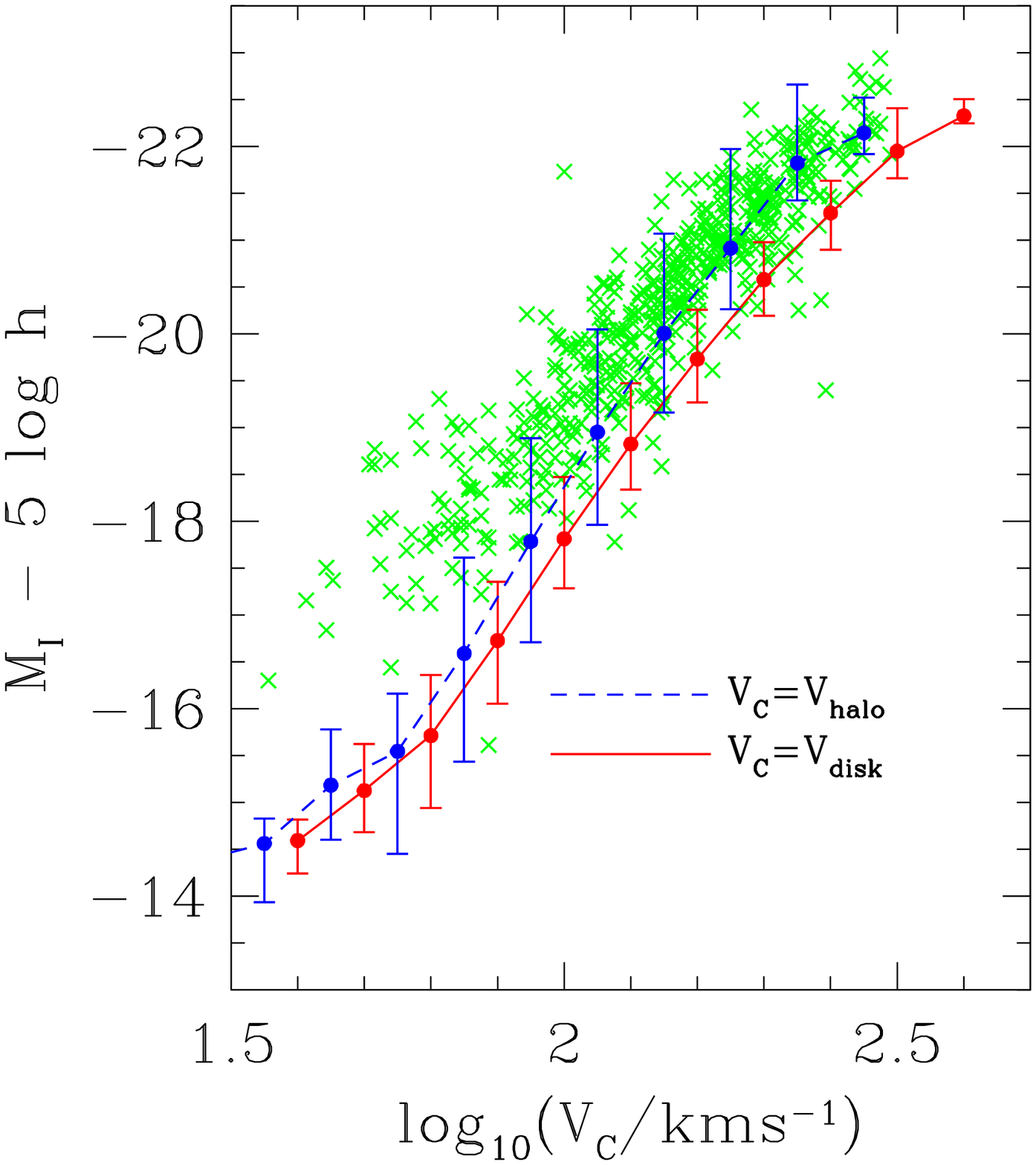}
\end{center}
\caption[]{
The Tully-Fisher relation for star forming disk galaxies. 
The crosses show data from the sample of Mathewson, Ford \& Buchhorn 
(1992). The dashed line shows the model prediction for the Tully-Fisher 
relation when the rotation speed of the halo at the virial radius is 
plotted. The solid line shows the predictions when the rotation speed 
at the half mass radius is plotted instead.
}
\label{fig:tf}
\end{figure}

In the phenomenological approach to galaxy formation, the values of the  
parameters in the recipes that describe processes 
such as star formation and feedback have to be specified to produce model 
predictions. This task is performed by comparing the model predictions 
to a subset of the available observational data. Different groups of 
modellers have different priorities when attempting to reproduce the data. 
The Munich group, for example, has attached most 
importance to matching the slope and zero point of Tully \& Fisher's (1977) 
correlation between the luminosity and rotation speed of disk 
dominated galaxies. The Durham group instead try hardest to 
match the form of the present day galaxy luminosity function. The 
luminosity function is the most basic description of the galaxy population 
and is now known to a high level of accuracy in the optical from the 
2dFGRS (Norberg \etal 2002) and SDSS (Blanton \etal 2000) and in the 
near-infrared from 2MASS photometry (Cole \etal 2001; Kochanek \etal 2001). 
The predictions of the fiducial model of Cole \etal (2000)  are compared 
with these recent estimates of the local luminosity function in Fig. 1.

Although most weight is given to reproducing 
the luminosity function when setting model 
parameters, matching other datasets, such as 
the Tully-Fisher relation, the distribution of disk scale lengths, 
the metallicity of gas in spiral disks and of stars in ellipticals, 
and the gas fraction in spiral disks, is also important. This greatly 
restricts the viable range of parameter space of the models.

One criticism levelled at semi-analytic models that has entered 
into popular folklore is the inability of the models to  
match the zeropoint of the Tully-Fisher relation at the same time 
as reproducing the break in the luminosity function at $L_{*}$.
The Tully-Fisher relation of the fiducial model of Cole \etal is 
compared with the observed relation in Fig. 2. The solid line shows 
the model prediction when the rotation speed at the half-mass radius 
of the disk is plotted; the dashed line shows how the zeropoint shifts 
when the rotation speed of the halo at the virial radius is plotted instead, 
which is much closer to the observed zeropoint. The shift is around 
20\% - 30\%, which is comparable to the accuracy one might expect in the  
calculation of the rotation speed at the half-mass radius. This depends 
upon several effects, such as the self gravity of the baryons and their 
gravitational influence on the halo dark matter.

\section{Model predictions - an example}

\begin{figure}
\begin{center}
\includegraphics[width=0.8\textwidth]{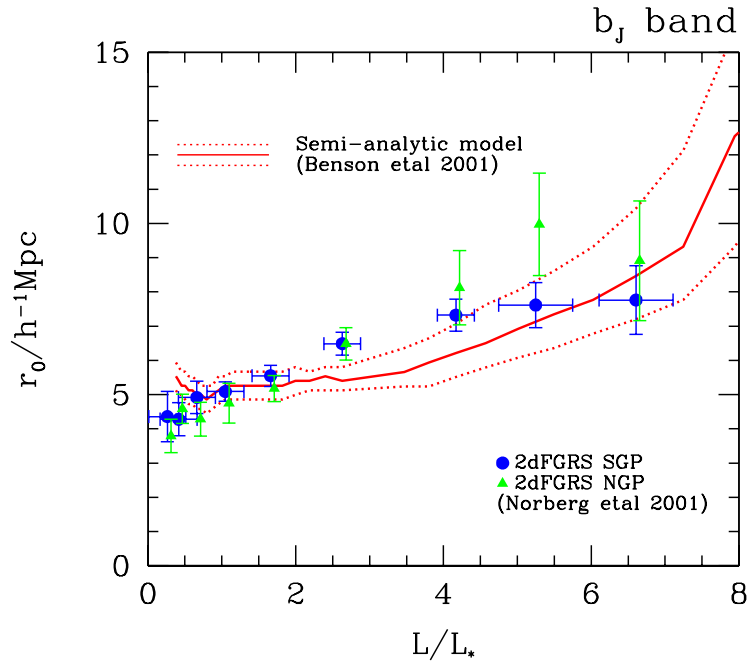}
\end{center}
\caption[]{The correlation length in real space, obtained by fitting 
a power law to the measured correlation function, 
$\xi(r) = (r_{0}/r)^{\gamma}$, plotted as a function of 
luminosity. The solid line shows the model predictions taken from 
Benson \etal (2001b). The dotted lines show the Poisson errors 
derived from the pair counts.
The symbols show the subsequent measurements made from the 2dFGRS 
(Norberg \etal 2001). In this case, the errors are estimated from 
mock 2dFGRS catalogues constructed from N-body simulations and include 
sample variance.
}
\label{fig:r0}
\end{figure}

Now that we have arrived at a fully specified model by comparing 
the output against a subset of the observations to fix the model 
parameters, we can make predictions for other quantities. 
Benson \etal (2000a,b; 2001b) populated a high resolution N-body 
simulation with galaxies using the semi-analytic model of Cole \etal. 
The simulation gives the spatial distribution of galaxies and allows 
their clustering to be measured. Remarkably, without any further 
adjustment to the model parameters, Benson \etal found that the fiducial 
$\Lambda$CDM model of Cole \etal predicts a correlation function 
that is in extremely good agreement with that measured for APM galaxies 
(Baugh 1996). This is particularly noteworthy as the galaxy correlation 
function is close to a power law, whereas the correlation function 
of the dark matter shows considerable curvature.

Benson \etal (2000b; 2001) presented predictions for the dependence 
of clustering strength on luminosity in the same model and found an 
approximately linear dependence of correlation length on luminosity; 
galaxies six times more luminous than $L_{*}$ have a correlation length 
$50\%$ longer than that predicted for $L_{*}$ galaxies. At the time, 
the picture emerging from the data was unclear. This has now been resolved 
by measurements from the 2dFGRS (Norberg \etal 2001) and SDSS (Zehavi \etal 
2002), which are in reasonable agreement with the trend predicted by  
the semi-analytic models. 

\section{The evolution of the stellar mass of galaxies}

\begin{figure}
\begin{center}
\includegraphics[width=0.9\textwidth]{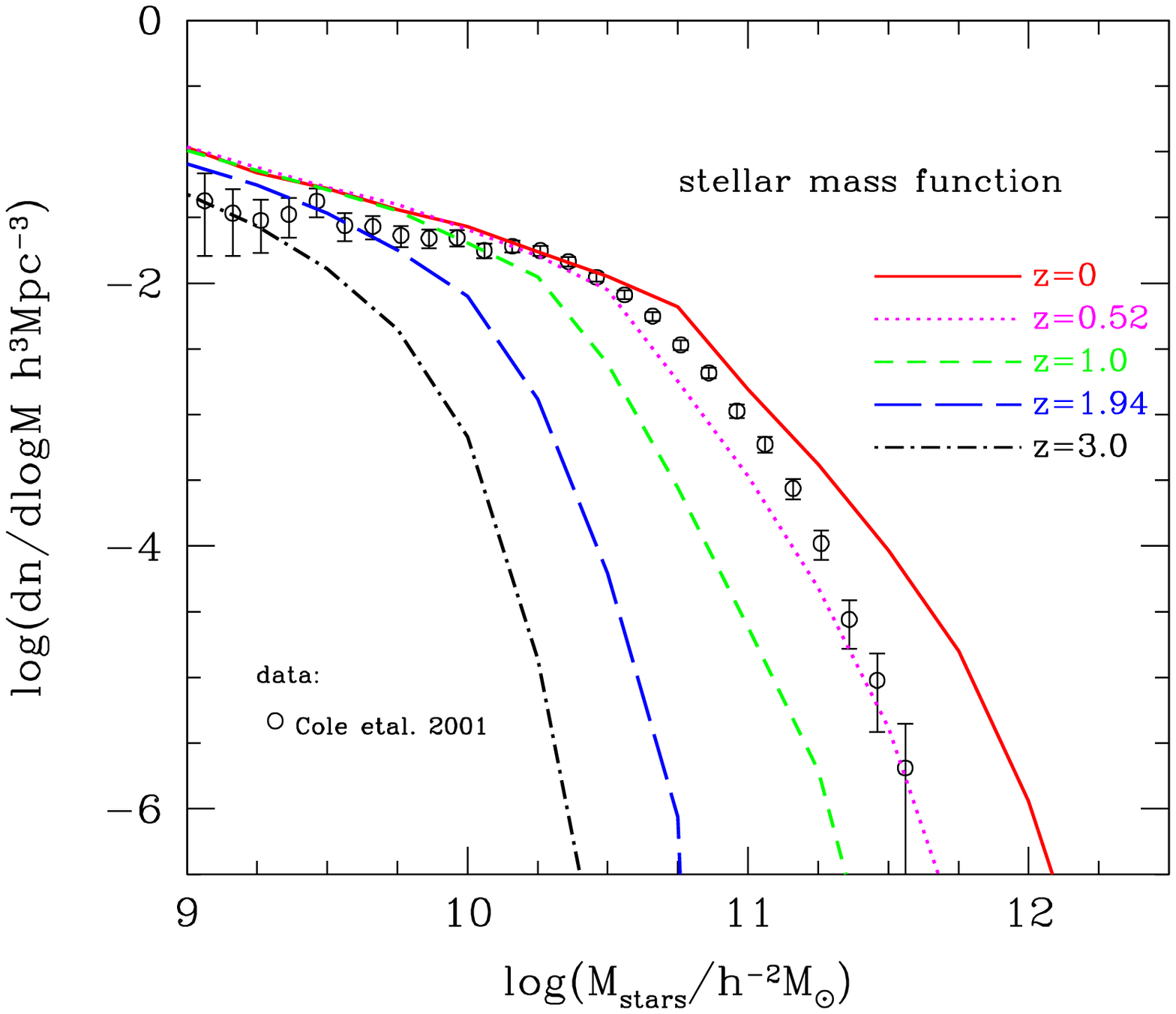}
\end{center}
\begin{center}
\includegraphics[width=0.9\textwidth]{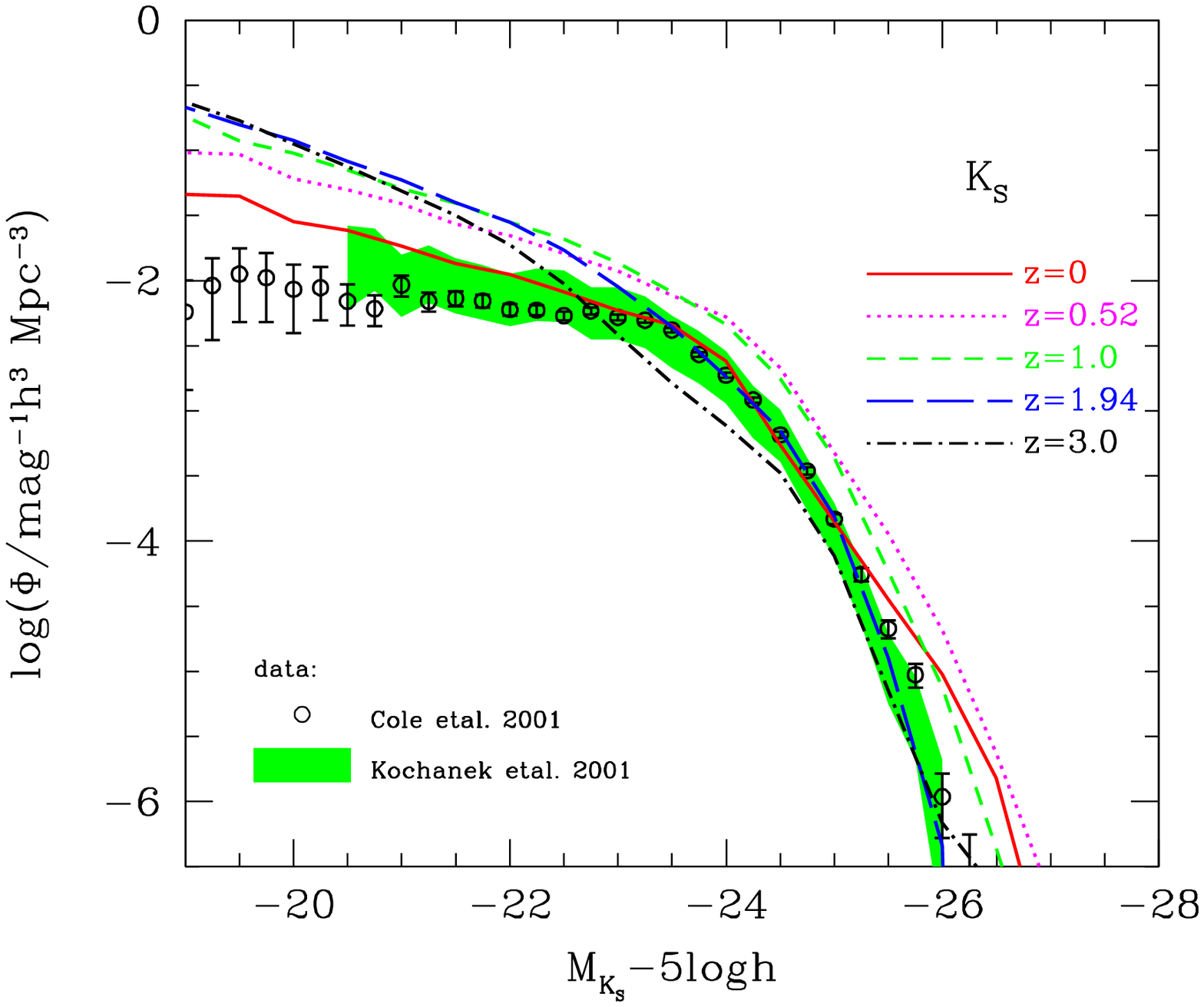}
\end{center}
\caption[]{
{\it Top:} The evolution of the stellar mass function with redshift. 
The lines show the model predictions at different redshifts, as indicated by 
the key. The datapoints show the present day stellar mass function 
inferred from the $K_{S}$-band luminosity function by Cole \etal (2001).
{\it Bottom:} The evolution with redshift of the {\it observer} 
frame $K_{S}$-band luminosity function. The symbols and shaded 
region show the present day $K_{S}$-band luminosity function 
estimated with 2MASS photometry.
}
\label{fig:mz}
\end{figure}

Advances in detectors that operate in the near infrared have led to 
a huge increase in the size of $K$ selected samples over the past decade. 
The first direct estimate of the $K$-band luminosity function 
from a $K$-selected 
sample used $\sim 500$ galaxies (Gardner \etal 1997); the estimate of 
the $K_{S}$-band luminosity function by Cole \etal (2001), using 2MASS 
photometry and 2dFGRS redshifts was made from over 17,000 galaxies.
The $K$-band luminosity of a galaxy gives a reasonable indication of its 
stellar mass. The output from the semi-analytic model suggests 
that the scatter in the stellar mass$--$$K$-band magnitude relation 
is a factor of $\sim 2$, showing the relative insensitivity to 
star formation history.

It is important to make a fair comparison between observational estimates 
and theoretical predictions for stellar mass. The stellar mass 
inferred from the $K$-band light is sensitive to the choice of IMF. 
Also, one needs to be clear whether recycling of gas is included i.e. 
whether the quantity under consideration is the mass locked up in stars or 
the total mass that had been turned into stars 
(some of which is subsequently expelled in stellar winds and supernovae). 
Cole \etal (2001) estimated the stellar mass function from the $K_{S}$-band 
luminosity function (shown by the symbols in Fig. 4), and found that only 
a small fraction of the baryons in the universe, perhaps as little as $5\%$, 
is actually locked up in stars. (Similar results were obtained by 
Kochanek \etal 2001.)

We plot the evolution of the stellar mass function in Fig. 4. 
There is a steady increase in the typical stellar mass 
with time; the value of $M_{*}$ increases by a factor of $\sim 2$ between 
$z=1$ and the present. The observable counterpart to the stellar mass 
function, the {\it observer} frame $K-$band luminosity function 
shows more complex evolution (Fig. 5). This is due to band shifting.

\vspace{0.2cm}

{\bf Summary} We have given an outline of the semi-analytic approach 
to modelling galaxy formation. This technique is complementary to 
direct simulation of the relevant gas dynamic processes. 
In fact, both methods rely upon physically motivated recipes to 
deal with star formation and feedback.  The model predicts strong 
evolution in the mass of stellar systems, with more than an order 
of magnitude increase in the abundance of $10^{11}h^{-2}M_{\odot}$ 
systems between $z=1$ and the present day.
Constraints on these predictions are now beginning to emerge, with 
the advent of the first results from deep,  near-infrared photometry 
(see, for example, Drory \etal 2001, and the contributions by 
Drory and by Papovich \etal to this volume).

%INDEX%%%%%%%%%%%%%%%%%%%%%%%%%%%%%%%%%%%%%%%%%%%%%%%%%%%%%%%%%%%%%%%
% Please check with the editor of your book whether he plans to
% include a "mutul" subject index - if so, please code your entries
% in the standard syntax. For your own purposes you may print your
% "personal" index by using the following commands:
%
%\clearpage
%\addcontentsline{toc}{section}{Index}
%\flushbottom
%\printindex
%%%%%%%%%%%%%%%%%%%%%%%%%%%%%%%%%%%%%%%%%%%%%%%%%%%%%%%%%%%%%%%%%%%%%

\end{document}